\documentclass[prb,twocolumn,showpacs]{revtex4}

\usepackage{graphicx}
\usepackage{amsmath}
\usepackage{amssymb}
\usepackage{epsf,latexsym}
\usepackage{times}
\usepackage{bm}
\newcommand{\ket}[1]{\left| #1 \right\rangle}
\newcommand{\bra}[1]{\left\langle #1 \right|}

\newcommand{\be}{\begin{equation}}
\newcommand{\ee}{\end{equation}}
\newcommand{\ba}{\begin{eqnarray}}
\newcommand{\ea}{\end{eqnarray}}

\begin{document}
\title{
Entanglement and density-functional theory: testing approximations on Hooke's atom.
}
\author{J. P. Coe$^{1,2}$
}
\email{jpc503@york.ac.uk}
\author{A. Sudbery$^{2}$
}
\email{as2@york.ac.uk}
\author{I. D'Amico$^{1}$
}
\email{ida500@york.ac.uk}
\affiliation{
$^1$ Department of Physics, University of York, York YO10 5DD, United Kingdom\\
$^2$ Department of Mathematics, University of York, York YO10 5DD, United Kingdom
}


\begin{abstract}

We present two methods of calculating the spatial entanglement of an interacting electron system within the framework of density-functional theory.  These methods are tested on the model system of Hooke's atom for which the spatial entanglement can be calculated exactly.  We analyse how the strength of the confining potential affects the spatial entanglement and how accurately the methods that we introduced reproduce the exact trends.  We also compare the results with the outcomes of standard first-order perturbation methods. The accuracies of energies and densities when using these methods are also considered.
\end{abstract}

\maketitle

\section{Introduction}

Entanglement, one of the stranger features of quantum mechanics, is now seen as a resource that can be exploited for teleportation of quantum states and secure distribution of cryptographic keys.\cite{NIELSEN}  In addition, it is thought of as one of the main reasons that quantum information devices may be able to outperform their classical counterparts.  In this respect many solid state systems have merit as quantum information processors, for example quantum dots.\cite{IDA1} However, modelling solid state many-body systems exactly is often computationally intractable, hence approximations are used.  One efficient method to model these systems is to use approximations within density-functional theory (DFT).\cite{DFTREVIEW}
DFT maps an interacting many-body system onto a non-interacting one, from which the ground state properties of the former can be calculated in principle exactly. In practice though, as the key quantity in DFT---the exchange-correlation potential---is an unknown functional of the density, approximations must be used.  
If we consider the entanglement of the ground state, then a relationship between DFT and the entanglement must exist;\cite{NOTE1,WU1,VIVIAN1} in this paper we propound ways to calculate the ground state entanglement of many-body systems using DFT.  We will consider the entanglement generated by the spatial degrees of freedom, and use Hooke's atom as a test system.  Hooke's atom represents a possible model for describing two electrons trapped in a quantum dot, a system similar to the ones proposed in various quantum information applications.  The interacting wave-function of this system can be calculated exactly.\cite{TAUT} This allows us to quantify its exact entanglement by calculating the Von Neumann and linear entropies.  We then propose two possible ways to calculate the same entanglement using DFT.  This is a delicate issue, as, by construction, DFT reproduces the exact density of the many-body system, but not its wave-function.
To calculate the entanglement, it is then necessary to define a suitable `interacting wave-function' within the DFT scheme.  
As a different route, we also explore the possibility of calculating the entanglement of this many-body system by means of perturbation theory.  We wish to discover how well these approximations reproduce the entanglement of the system and therefore ascertain whether they are good methods to model a system's suitability as a quantum information device (e.g. as a component for a quantum computer).

In Section II we introduce the system: Hooke's atom.  Section III discusses measures of continuous entanglement and uses them to quantify the exact entanglement for our system.  We also use the expectation values of the potential and Coulomb energies to express the degree of interaction as the confining potential varies.  In section IV we introduce DFT and use the local-density approximation (LDA) to calculate the approximate density and examine the link between the exact exchange-correlation energy and the entanglement.  The accuracy of the approximate entanglement when using the LDA is considered in section V where we define and calculate approximations to the `interacting LDA wave-function'.  In Section VI the second method to link DFT to an interacting wave-function is proposed.  This is based on perturbation theory, where the Kohn Sham equations are used as the zeroth-order Hamiltonian. We calculate the total energy, the density, and the entanglement to first-order in the perturbation using both the exact exchange-correlation potential and its LDA approximation; these results are then compared with first-order standard perturbation theory.  Finally, the paper concludes in Section VII with an overview of the different methods' ability to reproduce the exact entanglement and ideas for future work.

\section{The System}
The spatial entanglement of interacting electron systems has recently been investigated and calculated, by Osenda and Serra \cite{OSE2} using the Von Neumann entropy, for the spherical helium model where the Coulomb repulsion is replaced by its spherical average. Here we consider a system with full electron-electron interaction.  To do this we use the model system of two interacting electrons in a harmonic potential: Hooke's atom.  This system may be solved exactly \cite{TAUT} and  $2D$ and $3D$ harmonic potentials have been used to model particles trapped in quantum dots, for example see Refs.~\onlinecite{MERKT91,KUMAR90,CHOE98,IDA2,IDA3}.

In atomic units $\left (e=\hbar=m=1/4\pi\epsilon_{0}=1 \right)$ the Hamiltonian for two electrons confined by a harmonic potential is

\begin{equation}
\nonumber
H=-\frac{1}{2}\nabla^{2}_{1}-\frac{1}{2}\nabla^{2}_{2}+\frac{1}{2}\omega^{2}r_{1}^2+\frac{1}{2}\omega^{2}r_{2}^2+\frac{1}{\left|\bm{r_{1}}-\bm{r_{2}}\right|}
\end{equation}
where $\omega$ is the characteristic frequency of the confining potential.

This system has full electron-electron interaction yet may be separated into relative motion and centre of mass co-ordinates then solved exactly using the method of Taut.\cite{TAUT} This entails choosing $\omega$ so that the three term recurrence relation resulting from using a power series expansion for the solution of the relative motion radial part terminates. Unfortunately this means that exact ground-state solutions exist only for certain $\omega$, the largest of which is $0.5$.  For other values of $\omega$ we may construct an approximate solution to the relative motion part by using a finite number of terms ($400$) in the now infinite power series and numerically finding the eigenvalue of the relative motion part that causes its wave-function to tend to zero at large $r$. 

\section{Measures Of Entanglement And Exact Results}
Here we are considering fermionic entanglement.  For the two electrons the entanglement is spread over the spin and spatial parts of the wave-function.  We consider the ground state of the system so the spin part is that of a singlet and thus always maximally entangled, hence we concentrate on the spatial degrees of freedom.  The entanglement generated by the continuous spatial degrees of freedom will depend upon the interplay between the mutual electron repulsion and the strength of the confining potential.  Therefore, in this paper, we study how the spatial entanglement changes with the system parameters.  We shall quantify the entanglement using different measures i.e different entanglement entropies.  

\subsection{Linear entropy}

The linear entropy of the reduced density matrix
\begin{equation}
\nonumber
L=Tr \rho_{\text{red}}-Tr \rho_{\text{red}}^2=1-Tr \rho_{\text{red}}^2
\end{equation}
may be used as a measure of entanglement for a pure state (for example in Ref.~\onlinecite{WU1}) by giving an indication of the number and spread of terms in the Schmidt decomposition of the state.

We quantify the spatial entanglement of the electrons in our system by using the analogue of $L$ in the continuous case, where

\begin{equation}
\nonumber
\rho_{\text{red}}(\bm{r_{1}},\bm{r_{2}})=\int \Psi^{*}(\bm{r_{1}},\bm{r_{3}})\Psi(\bm{r_{2}},\bm{r_{3}})\bm{dr_{3}},
\end{equation}

\begin{equation}
\nonumber
\rho^{2}_{\text{red}}(\bm{r_{1}},\bm{r_{2}})=\int \rho_{\text{red}}(\bm{r_{1}},\bm{r_{3}})\rho_{\text{red}}(\bm{r_{3}},\bm{r_{2}})\bm{dr_{3}},
\end{equation}

\begin{equation}
\nonumber
Tr\rho^{2}_{\text{red}}=\int \rho^{2}_{\text{red}}(\bm{r},\bm{r})\bm{dr}. 
\end{equation}

\begin{figure}[ht]\centering
  \includegraphics[width=.4\textwidth]{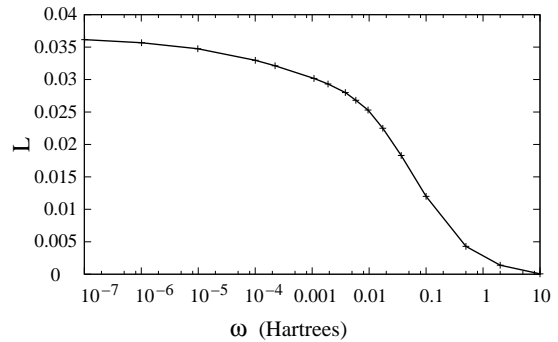}
  \caption{Linear entropy of the reduced density matrix versus $\omega$.}\label{fig:ExactL}
\end{figure}

The numerical calculations for varying $\omega$ are displayed in Fig.~\ref{fig:ExactL}.
Here we can observe, firstly, that the spatial entanglement is very low.  The linear entropy of the reduced density matrix is bounded by $d/(d-1)$, where $d$ is the dimension of the state.  As the spatial part is continuous then $d$ is infinite; hence the bound is unity, which our results are all very much less than.  The entanglement increases with decreasing confining potential as this means that the relative strength of the electron-electron interaction increases.  According to the value of $\omega$, we can identify three different regions.  For large values of $\omega$ the entanglement is essentially zero as we can neglect the electron-electron interaction compared to the confining potential.  In essence, in this region, we have two 3D isotropic oscillators hence the wave-function is a product in electron co-ordinates.  Another way to see this is to consider that as the confining potential tends to infinity then the electrons will be forced to overlap. This is allowed for a singlet state and the spatial part may be written as $\ket{0}\ket{0}$ which is clearly a product state.  For intermediate values of $\omega$, we see a sharp increase of the entanglement which corresponds to the confining potential and electron-electron interaction becoming of similar magnitude.  For $\omega \lesssim 0.001$ then the increase of entanglement per order of magnitude diminishes significantly.

We note that a quantum dot of width $2\lambda=10$nm corresponds to a confining potential of $\omega=3.65$ $\text{Hartrees}^{*}$ for GaAs and $\omega=0.684$ $\text{Hartrees}^{*}$ for CdSe.  Here $\text{Hartrees}^{*}$ are `effective Hartrees' calculated using the related material parameters and we have used the formula $\lambda=\sqrt{\frac{\hbar}{m^{*}\omega}}$.  The entanglement is therefore equivalent to the entanglement of Hooke's atom at $\omega=3.65$ Hartrees for GaAs and $\omega=0.684$ Hartrees for CdSe. 
\subsubsection{Interaction regions}
\begin{figure}[ht]\centering
  \includegraphics[width=.4\textwidth]{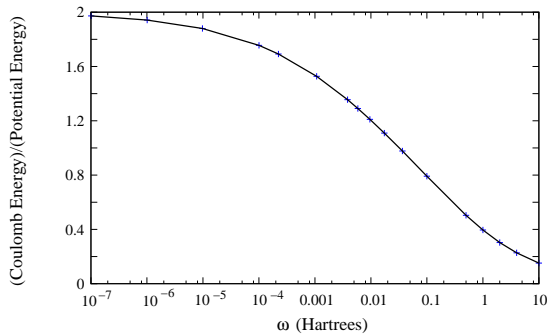}
  \caption{Ratio of the expectation of the Coulomb interaction to the expectation of the external potential versus $\omega$.}\label{fig:CoulombVpotential}
\end{figure}

To aid understanding of where the different regions of interaction occur we present a plot of the ratio of the expectation of the Coulomb interaction to the expectation of the external potential (Fig.~\ref{fig:CoulombVpotential}).  This allows us to identify three regions that help explain the behaviour of the entanglement in Fig.~\ref{fig:ExactL}: for $\omega\lesssim 0.001$ we have that the ratio is greater than $1.5$ and we deem this the `strongly interacting region' where the logarithmic plot of the entanglement appears to plateau;  $ 2 \gtrsim \omega \gtrsim 0.001$ corresponds to an `intermediate interaction region' where the logarithmic plot of the entanglement increases sharply; for $\omega \gtrsim 2$ the ratio shows that the external potential is the dominant term and we are entering the essentially non-interacting region where the entanglement is close to zero.

\subsection{Von Neumann entropy}\label{subsec-Von}
The Von Neumann entropy of the reduced density matrix is given by
\begin{equation}
\nonumber
S=-Tr \rho_{\text{red}} \log_{2}\rho_{\text{red}}=-\sum_{i}\lambda_{i}log_{2}\lambda_{i}, 
\end{equation}
and quantifies the entanglement for a bipartite pure state.  $S$ is thought of as the definitive measure of entanglement for a pure state (see for example Ref.~\onlinecite{BRUB02} and the references contained therein).  Here $\lambda_{i}$ are the eigenvalues of the reduced density matrix.

To calculate the continuous spatial entanglement, we first note that in the discrete case the eigenvalue equation may be written as
\begin{equation}
\nonumber
\sum_{j}\rho_{\text{red},ij}\phi_{j}=\lambda\phi_{i}.
\end{equation}
Hence in the continuous case we have
\begin{equation}
\int \rho_{\text{red}}(\bm{i},\bm{j})\phi(\bm{j})\bm{dj}=\lambda\phi(\bm{i}).
\label{eq:continS}
\end{equation}

This may be solved by turning it into an algebraic problem, following the idea of Osenda and Serra.\cite{OSE2}

We use a basis for the eigenfunction
\begin{equation}
\nonumber
\phi(r')=\sum_{n'}c_{n'}\eta_{n'}(r')
\end{equation}
and multiply Eq.~(\ref{eq:continS}) on the left by $\eta_{n}(r)$ and integrate to give the eigenvalue problem
\begin{equation}
\nonumber
\sum_{n'}\left (A_{nn'}-\lambda M_{nn'} \right )c_{n'}=0.
\end{equation}
Here
\begin{equation}
\nonumber
A_{nn'}=\int \eta_{n}(r)\rho^{red}(r,r') \eta_{n'}(r') dr dr'
\end{equation}
and
\begin{equation}
\nonumber
M_{nn'}=\int \eta_{n}(r) \eta_{n'}(r') dr dr'.
\end{equation}
To reduce calculation time and increase accuracy an orthonormal basis is created such that

\begin{equation}
\nonumber
4\pi\int_{0}^{\infty}\eta_{i}(r)\eta_{j}(r)r^{2}dr=\delta_{ij}.
\end{equation}

\begin{figure}[ht]\centering
  \includegraphics[width=.4\textwidth]{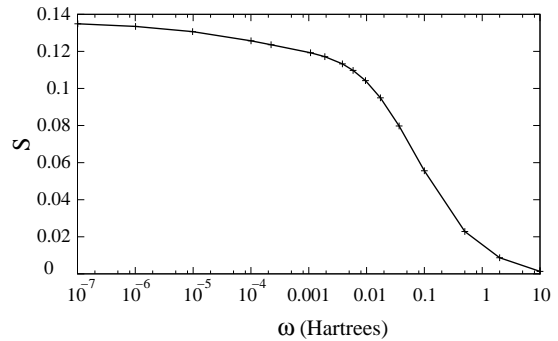}
  \caption{Von Neumann entropy of the reduced density matrix versus $\omega$.}\label{fig:ExactS}
\end{figure}

\begin{figure}[ht]\centering
  \includegraphics[width=.4\textwidth]{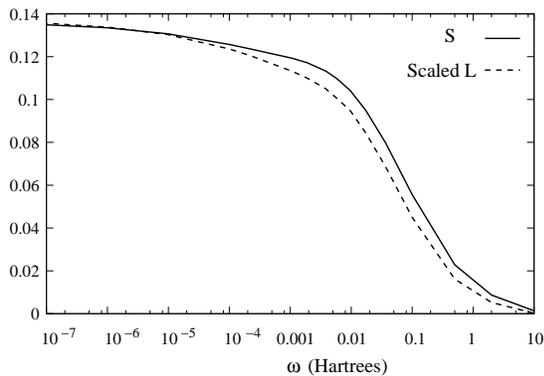}
  \caption{Comparison of $S$ and $L$ scaled by $3.75$ versus $\omega$.}\label{fig:Lscale}
\end{figure}

This is done by using the Gram-Schmidt process: we create polynomials $p_{i}(r)$ so that

\begin{equation}
\nonumber
\int_{0}^{\infty}p_{i}(r)p_{j}(r)w(r)dr=\delta_{ij},
\end{equation}

with the weight function chosen as $w(r)=r^{2}e^{-\omega_{r} r^{2}}$, where $\omega_{r}=\omega/2$.  We may then construct $\eta_{i}(r)=p_{i}(r)e^{-\omega_{r}r^{2}/2}$ and then normalise to give the orthonormal basis set.  The form of $\eta_{i}(r)$ is chosen to be similar to the exact solution. This process gives the results (Fig.~\ref{fig:ExactS}) which are of  a similar shape as the linear entropy (Fig.~\ref{fig:ExactL}) but of higher numerical values.  This is because for an infinite dimensional system, the Von Neumann entropy is unbounded unlike the linear entropy.
We compare the two measures of entanglement (Fig.~\ref{fig:Lscale}) by scaling $L$ by $3.75$ to achieve equality of the measures for the smallest value of $\omega$ plotted.  The measures do not scale exactly onto each other, but it is reassuring to see that they coincide in behaviour, at least for this system.  It appears that, for this confining potential range, the easier to calculate $L$, suitably scaled, is a good approximation to $S$.

\subsection{Position-space information entropy}

\begin{figure}[ht]\centering
  \includegraphics[width=.4\textwidth]{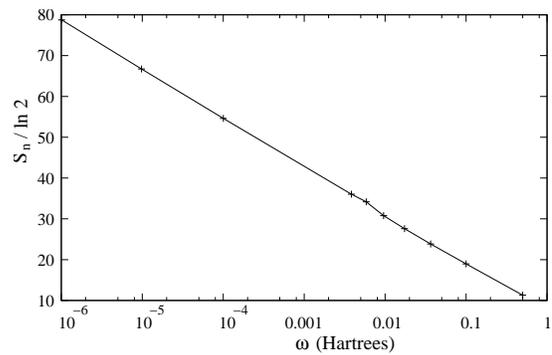}
  \caption{Position-space information entropy $S_{n}$ versus $\omega$.}\label{fig:ExactonlyapproxS}
\end{figure}

 The position-space information entropy $S_{n}$ has been used to study the Moshinsky atom\cite{AMOVILLI04} while the information entropy of the pair correlation function for Hooke's atom has also been studied by Atre et al.\cite{INFCOR} 
The position-space information entropy
\begin{equation}
S_{n}=-\int n(\bm{r})\ln n(\bm{r})\bm{dr}
\label{eqn:infentropy}
\end{equation}
may also be thought of as a zeroth-order approximation to $S$, where the off-diagonal terms of the reduced density matrix are set to zero.  Let us write
\begin{equation}
\nonumber
\rho_{red}=\rho_{\text{diag}}+\rho_{\text{off diag}}
\end{equation}
where
\begin{equation}
\nonumber
\rho_{\text{diag}}(r,r)=\frac{n(r)}{N},
\end{equation}
$n$ is the electron density and $N$ is the number of electrons.  Then, when neglecting off-diagonal terms we have
\begin{eqnarray*}
S \approx -\int \rho_{\text{diag}} \log_{2} \rho_{\text{diag}} dr \\
=-\frac{1}{N\ln 2}\int n(r) \ln n(r) dr +\frac{\ln N}{\ln 2}\\
=\frac{1}{N\ln 2}S_{n} +\frac{\ln N}{\ln 2}.
\end{eqnarray*}
Although much easier to calculate, our results in Fig.~\ref{fig:ExactonlyapproxS} show that the features of an exact measure of the entanglement are lost completely when using the position-space information entropy as an approximation to $S$.

\section{DFT treatment of Hooke's atom}

Solving the Schr\"{o}dinger equation directly for an interacting many-electron system is computationally onerous.  One framework that offers a more efficient method is density-functional theory, where the density rather than the many-body wave-function is employed to calculate properties of the system.  The theorem of Hohenberg and Kohn \cite{HK} states that for a non-degenerate ground state of an interacting electron system, the density uniquely determines the many-body wave-function.  Hence in theory all aspects of the ground state of the system are derivable from it.  The theory predicts a mapping between two systems having the same ground state density: the interacting system and the `Kohn Sham' system formed by {\it non-interacting} particles.  The density is found, exactly in principle, by solving single-electron equations describing the non-interacting system: the Kohn Sham (KS)\cite{KS} equations.  They contain the functional derivative of the exchange correlation energy functional $E_{\text{xc}}$ which allows us to move between the interacting and non-interacting systems.  

The KS equation for our system is
\begin{equation}
 \bigl(-\frac{1}{2}\nabla^{2}+\frac{1}{2}\omega^{2}r^{2}+v_{\text{H}}+v_{\text{xc}}(\bm{r};[n])\bigr)\phi(\bm{r})=\epsilon\phi(\bm{r}).
\label{eq:2eKS}
\end{equation}
Here
\begin{equation*}
v_{\text{xc}}=\frac{\delta E_{\text{xc}}}{\delta{n}}
\end{equation*}
is the exchange-correlation potential, and
\begin{equation}
v_{\text{H}}=\int \frac{n(\bm{r}_{2})}{|\bm{r}_{1}-\bm{r}_{2}|}\bm{dr_{2}}
\label{eq:vHartree}
\end{equation}
is the Hartree potential.

\subsection{Exchange-correlation energy as an entanglement indicator}

\begin{figure}[ht]\centering
\includegraphics[width=.4\textwidth]{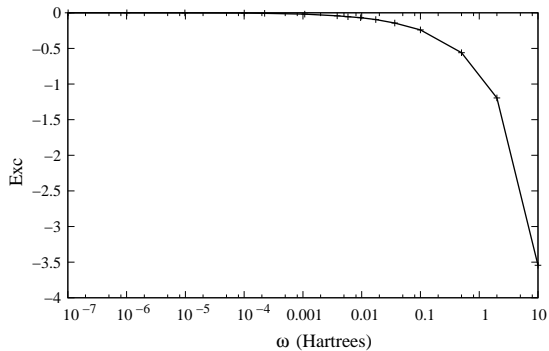}
  \caption{Exact $E_{\text{xc}}$ versus $\omega$.}\label{fig:logExc}
\end{figure}

Hooke's atom is one of the few systems where the exact $E_{\text{xc}}$ may be calculated by inversion of the KS equation. \cite{UMRIGAR}  We use this method to calculate the exact $E_{\text{xc}}$ for Hooke's atom for a range of $\omega$.  The results are depicted in Fig.~\ref{fig:logExc}.

\begin{figure}[ht]\centering
  \includegraphics[width=.4\textwidth]{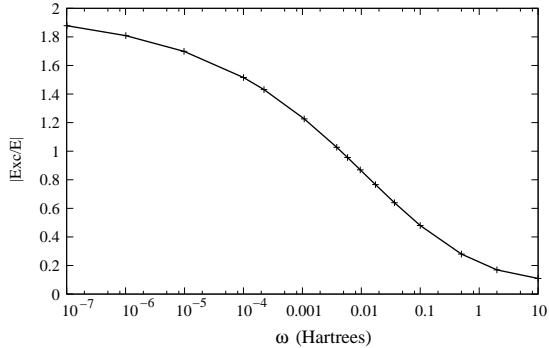}
  \caption{$|E_{\text{xc}}/E|$ versus $\omega$ as a possible entanglement indicator.}\label{fig:witness}
\end{figure}

As all the information about interactions is contained in $E_{\text{xc}}$, this would be expected to have a relationship with the entanglement.  It can be seen that $E_{\text{xc}}$ increases towards zero as $\omega$ decreases but appears to be unbounded as $\omega$ increases.  Hence, it bears little similarity to the entanglement behaviour.  This is because the interaction energy must be placed in context of the total system energy; therefore a better indicator would be $E_{\text{xc}}$ as a fraction of the energy of the system (Fig.~\ref{fig:witness}).  This follows a similar shape to the entanglement plot (Fig.~\ref{fig:ExactL}): it captures the general trend of the entanglement in the different interaction regions, but the increase in the `intermediate interaction region' is less sharp and the graph plateaus more slowly in the `strongly interacting region'.  Hence, we can think of this quantity as an indicator of the degree of entanglement for this system.

\subsection{LDA density results}
Unfortunately, the general functional form of $E_{\text{xc}}$ is unknown hence approximations are used.  One of the simplest, yet often effective, is that of $E_{\text{xc}}$ for a homogeneous electron gas, this is known as the local-density approximation (LDA).\cite{KS}

\begin{figure}[ht]\centering
  \includegraphics[width=.4\textwidth]{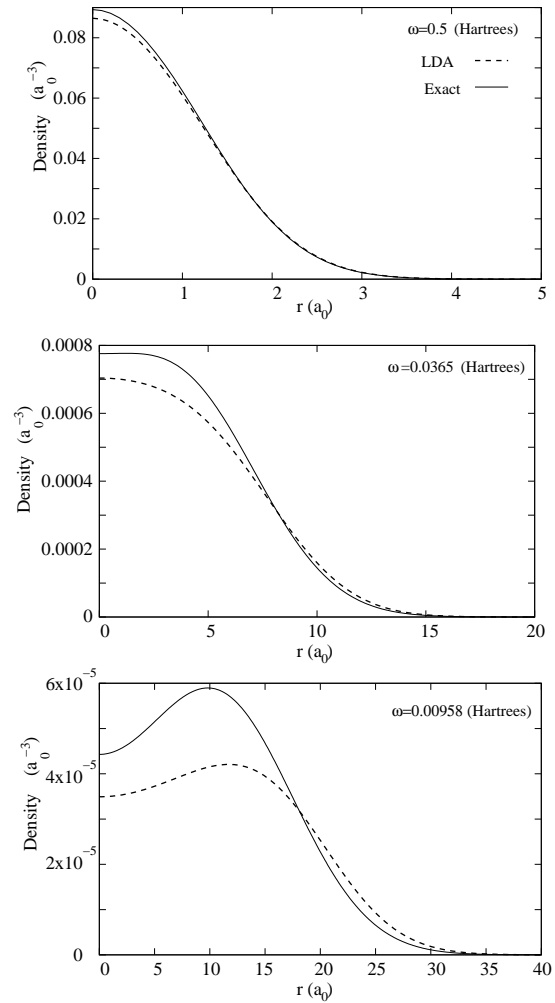}
  \caption{Comparison of the LDA density (dashed lines) and exact density (solid lines) plotted against r for three different values of $\omega$.}\label{fig:LDAcomparisonfusion}
\end{figure}

To aid understanding of how the LDA approximates entanglement we first investigate how well the LDA reproduces the effects of many-body interactions for Hooke's atom.  To achieve this we solve the KS equation (Eq.~(\ref{eq:2eKS})) with this approximation: for $v_{\text{xc}}$ we use $v^{\text{LDA}}_{\text{xc}}=v^{\text{LDA}}_{\text{x}}+v^{\text{LDA}}_{\text{c}}$ where $v^{\text{LDA}}_{\text{x}}$ is the exchange potential and $v^{\text{LDA}}_{\text{c}}$ is the correlation potential.  We employ the fit of Perdew and Wang \cite{PERDEW92} to Monte Carlo simulations for the correlation part while the expression for the exchange part is taken from Wigner and Seitz.\cite{WIGNER}
The KS equation is solved iteratively until self consistency is reached for the density $n(\bm{r})=2|\phi(\bm{r})|^{2}$.  Our solution is achieved by diagonalising the tri-diagonal matrix resulting from approximating the KS equation using central differences with the assumption that the ground state density is spherically symmetric.  In Fig.~\ref{fig:LDAcomparisonfusion} we present a comparison of $n_{\text{LDA}}(r)$, the self-consistent density resulting from using the LDA, and $n(r)$, the exact density, for three different values of $\omega$, as labelled. 

We quantify the relative distance between the LDA and exact density as 
  \begin{equation}
n~\%~\text{Error}=100\frac{\sum_{i}|n_{\text{LDA}}(r_{i})-n(r_{i})|}{\sum_{i}n(r_{i})}.
\label{eqn:npercenterror}
\end{equation}

\begin{figure}[ht]\centering
  \includegraphics[width=.4\textwidth]{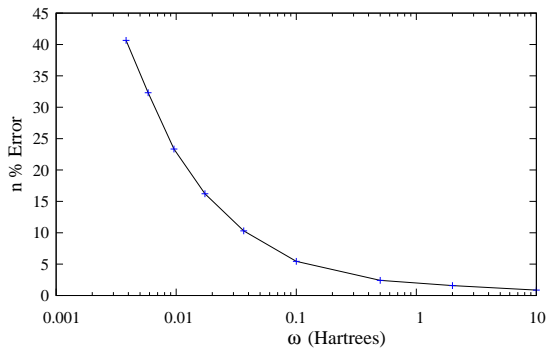}
  \caption{Relative error in the LDA density compared to the exact density versus $\omega$}\label{fig:LDAdensityerror}
\end{figure}

This is plotted in Fig.~\ref{fig:LDAdensityerror}. 

As $\omega$ decreases then the relative strength of the electron-electron interaction increases (see Fig.~\ref{fig:CoulombVpotential}).  We find that the LDA becomes less accurate as the relative strength of the many-body interactions in the system increases and that it tends to underestimate the density for small to medium $r$.  This is corroborated by the results of Taut, Ernst and Eschrig.\cite{TAUTDENSITY} This is somewhat counter-intuitive as with smaller $\omega$ the density is more slowly varying and hence more similar to that of a uniform electron gas.  It should be noted, however, that the self interaction, whereby the mean field approach of the Hartree potential Eq.~(\ref{eq:vHartree}) means that an electron is subject to an interaction with its own density, is not cancelled in the LDA approach.  This anomalous self interaction will be pronounced for this system as there are only two electrons.  The effect of the self interaction can be observed by the propensity of the LDA density to underestimate the exact density at the origin: the electrons feel more repulsion than they should so the density is spread further out.

\section{Entanglement in DFT: Method I\newline(The `Interacting LDA Wave-Function')}

DFT shows that all ground-state properties, therefore the entanglement, may be written as a functional of the ground-state single-particle density $n(\bm{r})$.  Unfortunately the explicit form of this functional for many properties, including the entanglement, is unknown.  However we know how to express the entanglement as a function of the interacting wave-function.  In this section we will propose a way of defining such a wave-function within the framework of DFT.   To test it we will calculate the corresponding entanglement of Hooke's atom and use the LDA to approximate the exchange-correlation potential.

We exploit the fact that the density uniquely determines the interacting wave-function, and search for the ground state interacting wave-function that reproduces the LDA density.

\subsection{Defining and finding an `interacting LDA wave-function'}
There are properties of the ground state of an interacting electron system which may be expressed easily in terms of the ground state wave-function but whose expression in terms of the density is unknown.  Two example of such properties are the kinetic energy and the entanglement.  The Hohenberg-Kohn theorem\cite{HK} showed that the ground state density for an interacting electron system uniquely determines its interacting wave-function.  Hence, knowing the ground-state density from DFT calculations, we can invert the problem and use the DFT density to find the ground-state interacting wave-function that reproduces the DFT density.  We will do this for Hooke's atom within the LDA and name the resultant wave-function as the `interacting LDA wave-function' $\psi_{\text{LDA,int}}$.  We may then calculate its entanglement. 

Our assumption is that if the LDA density closely approximates the exact density then $\psi_{\text{LDA,int}}$ will be close to the exact wave-function and consequently the same will be true of the entanglement.  The way we propose to achieve this mapping is to consider the LDA KS equations as {\it exact} KS equations for a fictitious interacting system, characterised by an external potential $v_{\text{ext}}^{\text{LDA}}$.  To circumvent the problem\cite{NOTE2} of the unknown exact form of $v_{\text{ext}}^{\text{LDA}}$, we choose then to design a method which reproduces the LDA density {\it and} minimises the functional 
\begin{equation}
Q=\bra{\Psi}\hat{T}+\hat{V}_{\text{ee}} \ket{\Psi} 
\label{eq:expect}
\end{equation}
 which appears in the KS theorem.  This ensures that the wave-function is indeed the ground-state as the method is equivalent to Levy's constrained search.\cite{LEVY} Here $\hat{T}$ is the kinetic energy operator and $\hat{V_{\text{ee}}}$ is the Coulomb energy operator.  Therefore, of the wave-functions which reproduce the density we choose the one that minimises Eq.~(\ref{eq:expect}): this is the wave-function that minimises the total energy of the system as the energy arising from the external potential, $\int v^{\text{LDA}}_{\text{ext}}(r)n_{\text{LDA}}(r) dr$, is fixed by the density.

Motivated by the form of the exact solution for Hooke's atom we choose the form of the wave-function we use in the search to be separable in centre of mass and relative motion co-ordinates

\begin{equation}
\psi_{\text{LDA,EA1}}(r,R)=p(r)e^{-\frac{1}{2}\omega_{r}r^{2}}e^{-2\omega_{R}R^{2}},
\label{eq:trialform}
\end{equation}
\begin{equation}
\nonumber
p(r)=\sum_{i}a_{i}r^{i}.
\end{equation}
Here $\omega_{r}$, $\omega_{R}$, and the $a_{i}$ are the parameters that we vary in the search and the subscript $EA1$  signifies evolutionary algorithm (see below). 
The $\psi_{\text{LDA,EA1}}(r,R)$ trial form for $\psi_{\text{LDA,int}}$ allows us to calculate the expectation Eq.~(\ref{eq:expect}) quickly.  Moreover, this method avoids solving the Schr\"{o}dinger equation each time or being overly restrictive in the form of $v_{\text{ext}}^{\text{LDA}}$.
To optimise the parameters we combine an evolutionary algorithm with a gradient descent algorithm to minimise the difference between the LDA density and the density $n_{\text{LDA,int}}$ arising from the trial interacting wave-function  
\begin{equation}
f=\frac{\sum_{i}|n_{\text{LDA}}(r_{i})-n_{\text{LDA,int}}(r_{i})|}{\sum_{i}n_{\text{LDA}}(r_{i})}.
\label{eqn:fitness1}
\end{equation}
  The outline of the first method to approximate the interacting wave-function is described in appendix A.
 We also consider the most general form of a two electron ground state wave-function arising from a spherically symmetric potential:
\begin{equation}
\psi_{\text{LDA,EA2}}(r_{1},r_{2},\theta)=\sum_{ijk}a_{ijk}\eta_{i}(r_{1})\eta_{j}(r_{2})P_{k}(\cos(\theta)),
\end{equation}
where
\begin{equation}
\eta_{i}(r)=p_{i}(r)e^{-\frac{1}{2}\omega r^{2}}
\end{equation}
are the functions orthonormal with respect to spherical polar integration detailed in Section \ref{subsec-Von} and we use $64$ terms.  In this method we consider the sum  of $f$ and $Q$ as the function to minimise.

\subsection{Approximation to the interacting LDA wave-function results}
\begin{figure}[ht]\centering
\includegraphics[width=.4\textwidth]{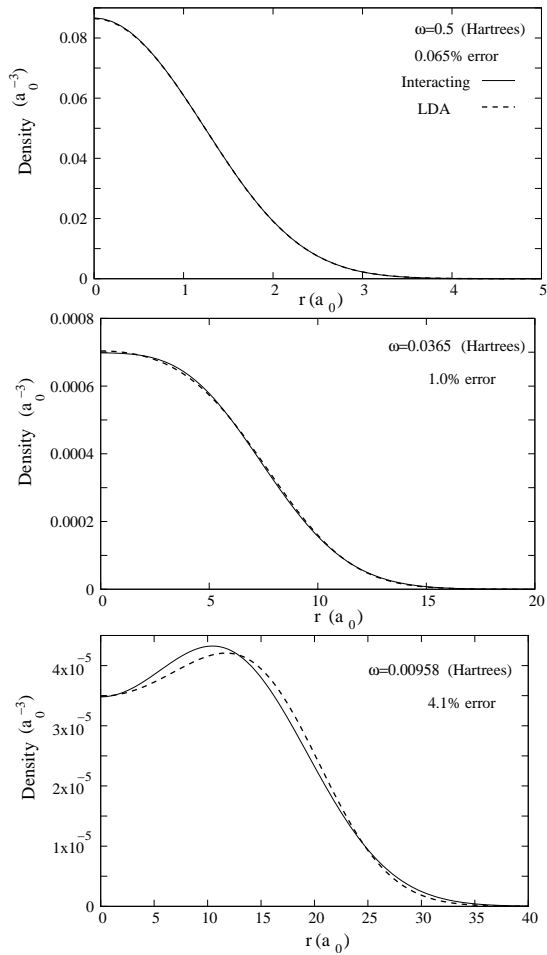}
 \caption{Comparison of the LDA density (dashed line) to the density arising from $\psi_{\text{LDA,EA1}}$ (solid line) plotted against $r$ for three different values of $\omega$.}\label{fig:GAcomparisonfusion}
\end{figure}

Fig.~\ref{fig:GAcomparisonfusion} compares $n_{\text{LDA}}(r)$ with $n_{\text{LDA,int}}(r)$, the density arising from $\psi_{\text{LDA,EA1}}$, for the best density match and lowest expectation Eq.~(\ref{eq:expect}) the algorithm found.  The error in match Eq.~(\ref{eqn:fitness1}) is displayed also (inset of Fig.~\ref{fig:GAcomparisonfusion}).
The match becomes less accurate as the importance of many-body interactions increases.  As we based this search around a wave-function of similar form to the exact solution then it is not surprising that the match is less accurate when the interactions increase as this is when the LDA density has a pronounced difference to the exact density.  It still should be noted that all these matches are much better than the match between the exact density and LDA density, compare Fig.~\ref{fig:LDAcomparisonfusion} and Fig.~\ref{fig:GAcomparisonfusion}.  We shall consider our approximate $\psi_{\text{LDA,int}}$ to be sufficiently accurate only when the relative error Eq.~(\ref{eqn:fitness1}) is less than a few percent. 

\subsubsection{Entanglement results}

\begin{figure}[ht]\centering
 \includegraphics[width=.4\textwidth]{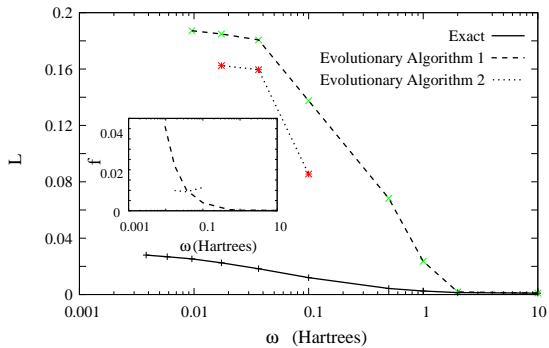}
 \caption{Comparison of $L$ for the exact wave-function and the approximate interacting wave-functions ($\psi_{\text{LDA,EA1}}$ and $\psi_{\text{LDA,EA2}}$) against $\omega$.  Inset: relative error in the density match $f$ versus $\omega$}\label{fig:LGAcomparison}
\end{figure}

The entanglement quantified using $L$ for the exact wave-function and the approximations to $\psi_{\text{LDA,int}}$ is depicted in Fig.~\ref{fig:LGAcomparison}, for values of $\omega$ that yielded a sufficiently accurate approximation to $\psi_{\text{LDA,int}}$; i.e $f \leq 4.1\%$ for $\psi_{\text{LDA,EA1}}$ and $f \leq 1\%$ for $\psi_{\text{LDA,EA2}}$ .  The next smallest $\omega$ for which an analytical exact solution to the Hooke's atom problem may be found led to $f \approx 8 \%$ hence was not included.  It can be seen that as $\omega\lesssim2$ (`intermediate strength' coupling regime) then the entanglement of our approximations to $\psi_{\text{LDA,int}}$ is greater than the true entanglement.  We find a large discrepancy between the entanglement when using the approximations to $\psi_{\text{LDA,int}}$ and the exact wave-function when the interaction strength becomes comparable to the confining potential see Fig.~\ref{fig:CoulombVpotential}.  Our calculations show that the entanglement calculated from this method is very sensitive to the trial form used for the wave-function and to the details of the fitness function ($f$ or $f+Q$) used.  
  
For larger $\omega$ the entanglement is reproduced.  We note that for $\omega=1$, $f$ is as low as $0.027\%$ and the entanglement is still overestimated using  $\psi_{\text{LDA,EA1}}$.  However when we consider $\psi_{\text{LDA,EA2}}$ for the intermediate regime the match is generally better and the overestimate is reduced.  Carrying out these calculations we realised that entanglement is very sensitive to small changes in the wave-function.  We have found approximations to $\psi_{\text{LDA,int}}$ that reproduce the $n_{\text{LDA}}$ accurately, have as low a energy as possible and are ground-states by virtue of their lack of nodes; however we can not guarantee that the energy is indeed the minimum.  The change in entanglement upon considering a more general basis set and so, in general, increasing the accuracy of the match, confirms for example that the energy from $\psi_{\text{LDA,EA1}}$ is indeed not the true ground state.  Due to this high sensitivity of the mapping between density and corresponding wave-function and of the entanglement on the wave-function details we can not be confident that we have found a close enough approximation to the exact $\psi_{\text{LDA,int}}$ to represent the entanglement trend correctly.  We speculate that a much larger number of iterations of the algorithm could improve accuracy, but this would become too computationally expensive to make this method a practical way to approximate the `interacting LDA wave-function'.

\subsection{Position-space information entropy}

\begin{figure}[ht]\centering
  \includegraphics[width=.4\textwidth]{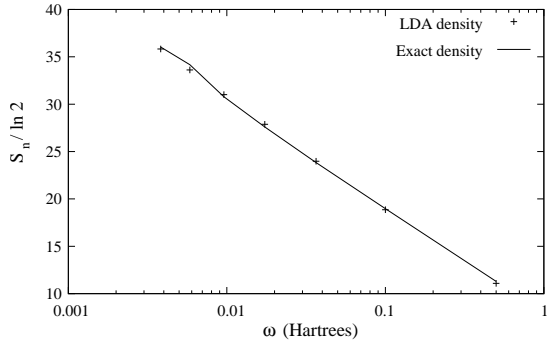}
  \caption{Position-space Information Entropy $S_{n}$ for the LDA density (crosses) and the exact density (solid line) plotted against $\omega$.}\label{fig:approxSgraph}
\end{figure}

  We return to the position-space information entropy $S_{n}$ Eq.~(\ref{eqn:infentropy}) and calculate it using $n_{\text{LDA}}$.  In Fig.~\ref{fig:approxSgraph} it can be observed that in this case the LDA gives almost the same information entropy as the exact solution.   
It is interesting that even when $n(r)$ and $n_{\text{LDA}}(r)$ are significantly different, i.e when $\omega$ is very small, the corresponding LDA and exact position-space information entropies are still very similar.  Clearly this measure is not sensitive to idiosyncrasies in the density shape.  $S_{n}$ is a relatively simple integral of a normalised quantity, the density.  As such, if two normalised functions are non-negligible in the same region and have roughly the same shape---as $n(r)$ and $n_{\text{LDA}}$ do (see Fig.~\ref{fig:LDAcomparisonfusion})---they will produce very similar values for $S_{\text{n}}$. This insensitivity is more evidence that $S_{n}$ is a poor approximation to an entanglement measure.

\section{Entanglement in DFT: Method II (Perturbation using the KS equations)}
The second method we propose to calculate entanglement within DFT, uses the KS equations as a zeroth-order Hamiltonian in a perturbation scheme chosen to give the exact Schr\"{o}dinger equation.  This scheme gives another method of associating an interacting wave-function with DFT.  

We write the full two-electron Hamiltonian for Hooke's atom as
\begin{equation}
H=H_{0}+H'
\label{eq:fullHam}
\end{equation}
where
\begin{equation}
H_{0}=\sum_{i=1,2}-\frac{1}{2}\nabla^{2}_{i}+\frac{1}{2}\omega^{2}r_{i}^2+v_{\text{xc}}\left(r_{i};[n] \right)+v_{\text{H}}\left(r_{i};[n] \right),
\label{eq:zeroorderHam}
\end{equation}
and
\begin{equation}
H'=\left(\sum_{i=1,2}-v_{\text{xc}}\left(r_{i};[n] \right)-v_{\text{H}}\left(r_{i};[n] \right)\right)+\frac{1}{|\bm{r}_{1}-\bm{r}_{2}|},
\label{eq:perturbedKSHam}
\end{equation}
is treated as the perturbative term.  Then the solution to Eq.~(\ref{eq:fullHam}) may be treated as a perturbation of the KS equations.  When $v_{\text{xc}}$ has to be approximated, $n(r)$ is found self-consistently for the ground-state and then used in the expression for $v_{\text{xc}}$ for the calculation of the excited states.  We expect this method to reduce the magnitude of the perturbation as many-body interactions are already partially accounted for in the zeroth-order term.  We therefore expect this to produce more accurate results in respect to a more standard perturbation expansion.  Similar procedures have been implemented \cite{GORLEVY1,GORLEVY2} to give an exact perturbation expansion for $E_{\text{c}}$.  In the following we will apply this method up to first order and examine how well it reproduces the entanglement of the system.  We will first perform our calculations by using the exact $v_{\text{xc}}$.  Hooke's atom is one of the few systems for which $v_{\text{xc}}$ may be expressed exactly by inverting the KS equations.\cite{UMRIGAR}  This corresponds to a test of the perturbation scheme only, as we have eliminated possible additional approximations in $v_{\text{xc}}$.

The details of the calculation of the first-order wave-function are given in Appendix B.  

\subsection{Comparison with standard perturbation theory}
We may also carry out a standard perturbation on the $3D$ harmonic oscillator equations.  Here we define
\begin{equation}
\nonumber
H=H_{0}+H'
\end{equation}
where
\begin{equation}
\nonumber
H_{0}=\sum_{i=1,2}-\frac{1}{2}\nabla^{2}_{i}+\frac{1}{2}\omega^{2}r_{i}^{2}
\end{equation}
and
\begin{equation}
\nonumber
H'=\frac{1}{\left|\bm{r}_{1}-\bm{r}_{2}\right|}.
\end{equation}

We use a similar method to that described in Appendix B to calculate the wave-function to first order.  We analytically calculate the total energy to first order as

\begin{equation}
\nonumber
E=E^{(0)}+E^{(1)}=3\omega+\sqrt{\frac{2\omega}{\pi}}.
\end{equation}

\subsection{Energy results}

\begin{figure}[ht]\centering
  \includegraphics[width=.4\textwidth]{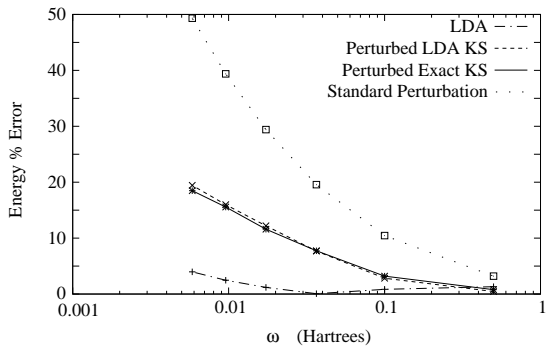}
  \caption{Relative error of the approximate energy from LDA, first-order perturbation of the LDA KS equations, first-order perturbation of the exact KS equations and standard first-order perturbation, compared to the exact energy plotted against $\omega$.}\label{fig:Energycomparison}
\end{figure}

\begin{figure}[ht]\centering
  \includegraphics[width=.4\textwidth]{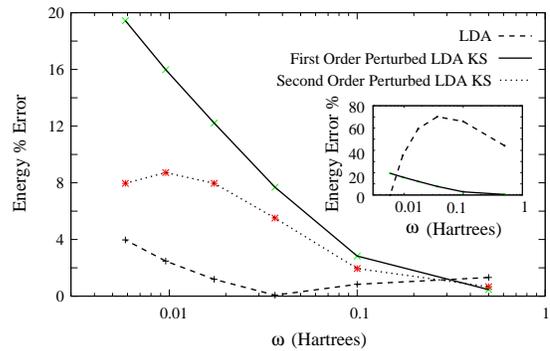}
  \caption{Comparison of the relative errors of first and second-order perturbations of the LDA KS equations together with the relative error in the LDA energy plotted against $\omega$.  Inset: Comparison of the relative errors of zeroth (dashed line) and first-order (solid line) perturbations of the LDA KS equations compared to the exact energy plotted against $\omega$. }\label{fig:EnergycomparisonwithE2}
\end{figure}

We first compare the results for the total energy of the system. In Fig.~\ref{fig:Energycomparison} we plot the relative error
\begin{equation}
\text{Energy}~\%~\text{Error}= 100 \left( \frac{E_{\text{appx}}-E}{E} \right),
\end{equation}
where $E$ is the exact system energy and $E_{\text{appx}}$ is the system energy calculated using an approximation: the LDA; first-order perturbation of the exact KS equations; first-order perturbation of the LDA KS equations; or first-order standard perturbation.  It can be seen that standard first-order perturbation fares poorly in calculating the energy, while this is greatly improved by the perturbation of the LDA KS equations to first order but only marginally improved again by the perturbation of the exact KS equations.  The LDA energy though is the most accurate and performs very well with an error always less than $6\%$ for the plotted confining potentials.  It is interesting that the LDA reproduces the energy much better for weak confining potentials than it reproduces the density or entanglement. 
 As an additional check we may calculate the energy to second order for the perturbation of the LDA KS equations (Fig.~\ref{fig:EnergycomparisonwithE2}).  This can be seen to reduce the error, especially for the low confining potentials, although it still does not achieve as high an accuracy as the LDA energy.   In the inset of Fig.~\ref{fig:EnergycomparisonwithE2} we show that when implementing a perturbation of the LDA KS equations the first-order energy is generally more accurate than the zeroth-order energy.  However, the non-trivial behaviour of $E^{(0)}$ (twice the KS eigenvalue) results in propitious closeness of $E^{(0)}$ to the exact energy for $\omega\approx 0.00584$.
\subsection{Entanglement results}
\begin{figure}[ht]\centering
  \includegraphics[width=.4\textwidth]{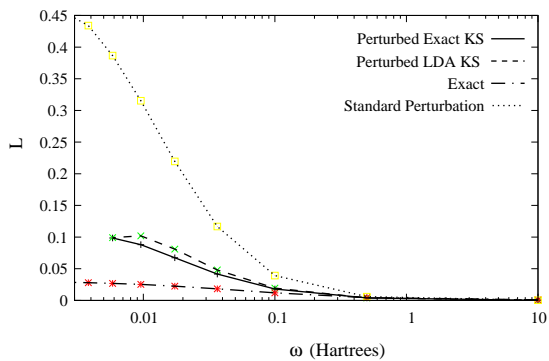} 
  \caption{Entanglement quantified with $L$ of the exact solution, first-order perturbation of the KS equations, first-order perturbation of the LDA KS equations and standard first-order perturbation, plotted against $\omega$.}\label{fig:LperturbationstandardandLDA}
\end{figure}

Fig.~\ref{fig:LperturbationstandardandLDA} shows that standard perturbation approximates the entanglement well in the essentially non-interacting region but poorly for confining potentials smaller than $\omega \approx 0.5$.  The perturbation of the KS equations to first order performs much better both for the exact $v_{\text{xc}}$ and its approximation $v_{\text{xc}}^{\text{LDA}}$.  It reproduces the entanglement well for $\omega\gtrsim 0.5$ after which it begins to overestimate the entanglement.  This overestimate becomes worse as $\omega$ decreases, but the accuracy is still superior to that of standard perturbation.  An explanation for the behaviour of the approximations to the entanglement shown in Fig.~\ref{fig:LperturbationstandardandLDA} is that at strong confining potentials the perturbation is very small and hence first-order perturbation will give an accurate wave-function with the correct entanglement both for the standard perturbation and perturbation of the KS equations.  When the interaction increases perturbation theory is bound to become less accurate.  In particular the resulting first-order wave-function will be mainly composed of a mixture of excited state wave-functions, therefore the entanglement will increase.  We find that this mixture consistently overestimates the entanglement. 

 Our hypothesis was that the perturbation term $H'$ (Eq.~(\ref{eq:perturbedKSHam})), when using the KS equations as the zeroth-order Hamiltonian, was smaller than the standard perturbation term, due to some of the interaction being taken into account in the KS equations, which represent, in this case, our zeroth-order Hamiltonian.  This implies that the related entanglement will be more accurate, which is supported by our results.  To investigate the accuracy of the first-order perturbation of the KS equations when $v_{\text{xc}}$ is approximated we also consider this perturbation using $v_{\text{xc}}^{\text{LDA}}$.  We see that the entanglement calculated using perturbation of the LDA KS equations is very similar to that calculated with perturbation of the exact KS equations, although there is a slight deterioration in the accuracy. Therefore the LDA may be thought of as sufficiently accurate for this method and little is gained by using the exact exchange-correlation potential.  Hence for smaller confining potentials ($\omega \lesssim 0.5$) it seems that higher-order perturbation is required to improve the accuracy of the calculated entanglement, not a more accurate exchange-correlation potential, at least for this system.  Around $\omega\approx 0.01$ the perturbed LDA KS entanglement starts to decrease slightly; after this it appears that the numerical accuracy becomes insufficient.  The range we can accurately perform the calculation on is similar to the range we were successful in approximating $\psi_{\text{LDA,int}}$, hence it allows us to compare the two methods.

\begin{figure}[ht]\centering
  \includegraphics[width=.4\textwidth]{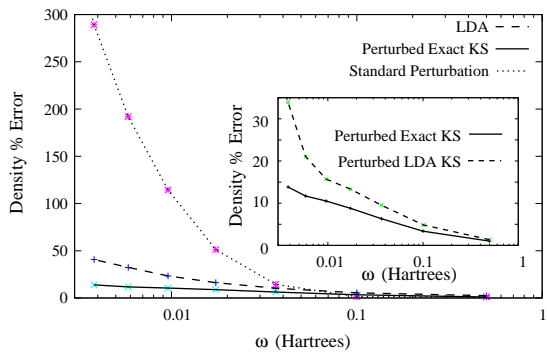}
  \caption{Comparison of the relative error in the reproduction of the density by approximate methods versus $\omega$.  Inset: comparison of the relative error in the reproduction of the density between perturbation of the exact KS equations and perturbation of the LDA KS equations versus $\omega$.}
 \label{fig:Densityerrorcomparisongraph}
\end{figure}

We may also quantify the accuracy of the method by comparing the match between the density calculated using the various approximations and the exact density, using an expression similar to Eq.~(\ref{eqn:npercenterror}). As can be seen in Fig.~\ref{fig:Densityerrorcomparisongraph} the perturbation using the exact KS performs slightly better than $n_{LDA}$ so reversing the result of the energies.  The standard perturbation is poor and becomes rapidly worse as $\omega$ decreases. We also compare the perturbation of the exact KS equations to the perturbation of the LDA KS regarding the accuracy of the density in the inset of Fig.~\ref{fig:Densityerrorcomparisongraph}.  As with the entanglement there is an improvement when using the exact $v_{\text{xc}}$, especially for small $\omega$, but the increase in accuracy is much larger for the density.

\section{Overview and Conclusions}
We have calculated the entanglement for the spatial degrees of freedom of two electrons confined by a harmonic potential (Hooke's atom) and compared different measures of the entanglement for this system.  We found that the linear and Von Neumann entropy of the reduced density matrix give consistently similar results: for Hooke's atom in its ground state the entanglement increases with increasing electron-electron interaction but is always quite low.  The behaviour of the entanglement, in this case, is therefore independent of which of these exact measures is employed. We have also studied the position-space information entropy, which does not produce the correct behaviour of entanglement when the strength of the external potential is varied.  As the position-space information entropy can be viewed as a simple and appealing approximation to the Von Neumann entropy we conclude that caution must be used when dealing with it; more analysis is necessary to understand for which systems such an approximation is valid.  We analysed a possible entanglement indicator, the exchange-correlation energy $E_{\text{xc}}$.  We found that despite all the many-body interactions being contained within this quantity it does not reproduce the features of the entanglement. We calculated $|E_{\text{xc}}/E|$ and found it a better indicator of entanglement: it gave a similar trend to that of $S$ or $L$ with respect to the confining potential.

In the second part of the paper we introduced two possible ways of calculating the entanglement of a many-body system within density-functional theory.  This is a powerful method to treat many-body systems, but there is no obvious way within it to calculate the system wave-function, which we needed to deduce the entanglement.  To do so we have proposed an `interacting LDA wave-function' $\psi_{\text{LDA,int}}$ and a perturbation scheme based on the KS equations.  We have applied these methods to the Hooke's atom problem, the first using the LDA to approximate the exchange-correlation potential $v_{\text{xc}}$ and the second using the exact $v_{\text{xc}}$ and its LDA approximation.  

As a precursor to understanding the entanglement we looked at the accuracy of the density reproduction when using the LDA.  We found that when using the LDA to model this system the resultant density is very close to the exact density when the confining potential is large, however the match becomes progressively worse as the relative strength of the electron-electron interaction increases.  

\begin{figure}[ht]\centering
  \includegraphics[width=.4\textwidth]{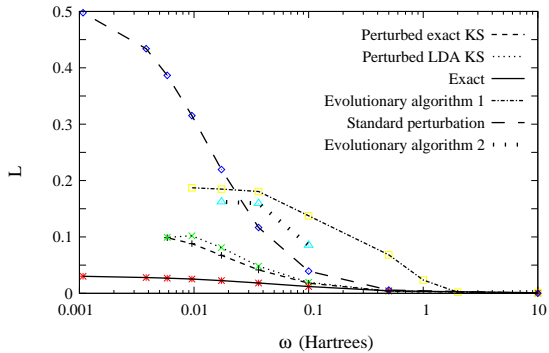}
  \caption{Overview of L for the approximate methods and exact solution plotted against $\omega$.}\label{fig:Loverviewgraph}
\end{figure}

\begin{figure}[ht]\centering
  \includegraphics[width=.4\textwidth]{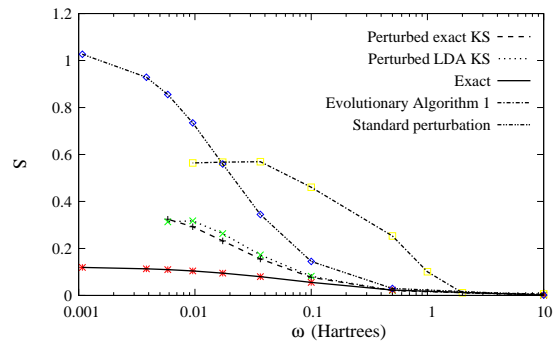}
  \caption{Overview of S for the approximate methods and exact solution plotted against $\omega$.}\label{fig:Soverviewgraph}
\end{figure}

To reproduce the entanglement, we defined an `interacting LDA wave-function' $\psi_{\text{LDA,int}}$, which reproduces the LDA density and minimises the expectation of the kinetic and Coulomb energies.  As an approximation to $\psi_{\text{LDA,int}}$, $\psi_{\text{LDA,EA1}}$ and $\psi_{\text{LDA,EA2}}$ were found using a combined evolutionary and steepest descent algorithm.  These approximations were used to calculate the entanglement and revealed that they overestimate the entanglement when there is non-negligible electron-electron interaction ($\omega \lesssim 2$) (Fig.~\ref{fig:Loverviewgraph} and Fig.~\ref{fig:Soverviewgraph}).  However we found a discrepancy between the entanglement calculated using different trial functions even for very accurate density matching.  This suggests that the mapping between the density and the interacting wave-function in a DFT scheme is highly non-linear and confirms the sensitivity of the entanglement to the details of the wave-function.  As a result we cannot know whether the ground-state we have found is close enough to that of the interacting LDA system, so this method to find the `interacting LDA wave-function' does not work well.  We speculate that the search for the interacting LDA wave-function would be more successful if it were possible to search generally and efficiently for the external potential of an interacting electron system that gives the LDA density.

 We then considered perturbative methods.  We first analysed the behaviour of the system total energy.  The results showed that standard first-order perturbation is inaccurate but is much improved by using the exact KS equations as a zeroth-order Hamiltonian, with only a small decrease in accuracy when the LDA KS equations are used instead.  The LDA energies are the most accurate.  This shows how errors are cancelled in the LDA to give accurate energies despite the density and entanglement being inaccurate for strong interactions.  For the accuracy of the density reproduction we observed that standard first-order perturbation performed worst and a large improvement was given by perturbing the LDA KS equations.  This gave an even higher accuracy than the LDA density.  Density reproduction was further improved (markedly so for small $\omega$) by using the KS equations with the exact $v_{\text{xc}}$.  

The results obtained for the entanglement are summarised in Fig.~\ref{fig:Loverviewgraph} and Fig.~\ref{fig:Soverviewgraph}.  The entanglement calculated using first-order perturbation of the LDA KS equations was larger than the exact entanglement but not until after $\omega \approx 0.5$ so this is already an improvement on the results from the approximations to $\psi_{\text{LDA,int}}$.  In addition, when it failed to accurately approximate the entanglement, the overestimate was less severe than when using $\psi_{\text{LDA,EA1}}$ or $\psi_{\text{LDA,EA2}}$. The accuracy of the entanglement when using perturbation of the KS equations was only slightly improved by using the exact $v_{xc}$.  This suggested that higher-order terms are required, not more accurate exchange-correlation energy functionals, when employing this method.  Standard perturbation was again the least accurate amongst the perturbative methods, severely overestimating the entanglement when the confining potential was small.  This supported the hypothesis that a perturbation of the KS equations would approximate the exact results more accurately than standard perturbation as some of the interaction is already included in the KS equations.  Even though far from optimal, standard perturbation is still an improvement over the results for the approximations to $\psi_{\text{LDA,int}}$ for $\omega \gtrsim 0.03$.  We also note that these trends are observed whether we use the linear entropy of the reduced density matrix $L$ or the Von Neumann entropy of the reduced density matrix $S$, as can be seen by comparing Fig.~\ref{fig:Loverviewgraph} and Fig.~\ref{fig:Soverviewgraph}.  

We find that all our methods to approximate $L$ and $S$ overestimate the entanglement.  For the approximations to the entanglement involving the LDA,  we suggest that, in addition to the discussed limitations of the methods, it is also an effect of the self-interaction (see discussion of Fig.~\ref{fig:LDAcomparisonfusion}), which effectively enhances the interaction and hence the entanglement.  It is less clear, though, why the entanglement is overestimated, rather than underestimated, when using perturbation of the exact KS equations and standard perturbation.  

The results have indicated that neither our approximation to $\psi_{\text{LDA,int}}$ nor first-order perturbation approximate the spatial entanglement accurately when considering strongly interacting systems, yet first-order perturbation give the energy to a relatively high accuracy.  Hence it seems that entanglement is much more susceptible to small changes in the density than the energy: the entanglement is an integrated quantity of a wave-function, which is very sensitive to changes in the density.  

The issue of calculating an entanglement within the LDA has also been explored by V. V. Fran\c{c}a and K. Capelle,\cite{VIVIAN1} but in the context of the Hubbard model and spin entanglement.  We concur with their conclusion that spatial inhomogeneity (in our case determined by the strength of the confining potential) has a major effect upon entanglement.

 Further work is planned to investigate whether we can efficiently find the external potential of the interacting system that reproduces the LDA density: the ground-state of such a system would be the true `LDA interacting-wave-function' $\psi_{\text{LDA,int}}$.  We then intend to investigate whether applying a self-interaction correction (SIC) to the LDA will improve the approximation of the density: the self interaction becomes a predominant error as the number of electrons decrease, hence by quelling this anomalous effect to some degree in Hooke's atom we would hope to make improvements in density accuracy and therefore the approximation of the entanglement. 

\section{Acknowledgements}
We are grateful to K. Capelle for valuable and fruitful discussions.  I. D. and J. P. C. acknowledge the kind hospitality of the Department of Physics, University of S\~{a}o Paulo, S\~{a}o Carlos campus.  J. P. C. wishes to thank the EPSRC for providing financial support.

\appendix
\section{Combined evolutionary and gradient descent algorithm}
The following is the layout of the computational method we used to calculate $\psi_{\text{LDA,int}}$.
\begin{itemize}
  \item Create a collection of wave-functions with random parameters (polynomial coefficients and  $\omega_{r}$, $\omega_{R}$).
  \item Minimise the fitness function (Eq.~(\ref{eqn:fitness1})) by using a gradient descent algorithm on the parameters of each wave-function
  \item Keep the closest matches and create other wave-functions by randomly mutating parameters in the closest matches and creating new wave-functions with random parameters.     
  \item The previous two steps are then repeated.
\item  After around thirty generations the program ends.  Of the overall best ten matches, those with the lowest expectations of the kinetic and Coulomb energy (Eq.~(\ref{eq:expect})) are chosen as the best approximations to the interacting ground-state wave-function.
\end{itemize}

To ensure that we have found a ground-state wave-function the ten best matches are also checked to verify that they are node-less and that they are lower in energy than the non-interacting wave-function formed by the product of the single electron KS orbitals.

\section{First-order perturbation using the KS equations}
We require the ground state wave-function to first-order 
\begin{equation}
\nonumber
\Psi_{0}\approx \Psi_{0}^{(0)}+\Psi_{0}^{(1)}.
\end{equation}
To find $\Psi_{n}^{(1)}$ we use the Rayleigh-Schr\"{o}dinger method, where the first-order wave-function is written in terms of the unperturbed basis

\begin{equation}
\Psi_{n}^{(1)}=\sum_{k=0}^{\infty} a_{nk}^{(1)} \Psi_{k}^{(0)}. 
\label{eq:appUnperturbedbasis}
\end{equation}

Here
\begin{equation}
a^{(1)}_{nk}=\frac{1}{E_{n}^{(0)}-E_{k}^{(0)}}\bra{\Psi_{k}^{(0)}}H'\ket{\Psi_{n}^{(0)}}.
\label{eq:appCoeffs}
\end{equation}

and $a_{nn}^{(1)}=0$ (normalisation of the first-order wave-function).  Here we concern ourselves with the ground state ($n=0$) only.
We solve 
\begin{equation}
\nonumber
H_{0}\Psi_{n}^{(0)}=E_{n}^{(0)}\Psi_{n}^{(0)}
\end{equation}
by separating $H_{0}$ for each electron, then separating into radial and angular parts and solving the radial part using exact diagonalisation techniques.  

We write a single electron wave-function as
\begin{equation}
\nonumber
\phi_{nlm}(\bm{r})=R_{nl}(r)Y_{lm}(\theta,\phi).
\end{equation}

Using the ansatz $R_{nl}(r)=u_{nl}(r)/r$ gives the radial equation

\begin{eqnarray}
\nonumber&~&\Bigl(-\frac{1}{2}\frac{d^{2}}{dr^{2}}+\frac{1}{2}\omega^{2}r^{2}+v_{\text{xc}}+v_{\text{H}}\\
\nonumber&~&+\frac{l(l+1)}{2r^{2}}\Bigr)u_{nl}(r)=\epsilon_{nl}u_{nl}(r).
\end{eqnarray}

 For $l=0$ this is solved self-consistently when $v_{\text{xc}}$  has to be approximated, then the ground-state self-consistent density is used in the calculation of $v_{\text{xc}}$ for the higher angular momentum states.  From the eigenvalues there appear to be no degeneracies due to the non-trivial nature of $v_{\text{xc}}$ and $v_{\text{H}}$.

The exact wave-function is a singlet, therefore any triplet states will have zero contribution or will cancel out; hence in our scheme we chose to neglect their contribution in Eq.~(\ref{eq:appUnperturbedbasis}). The spin singlet excited state wave-functions can be divided into two forms:

\begin{itemize}
\item{Product states}
\begin{equation}
\nonumber
\phi_{n_{1}l_{1}m_{1}}(\bm{r_{1}})\phi_{n_{2}l_{2}m_{2}}(\bm{r_{2}});
\end{equation}

\item{Non-product states}
\begin{equation}
\frac{1}{\sqrt{2}}\left(\phi_{n_{1}l_{1}m_{1}}(\bm{r_{1}})\phi_{n_{2}l_{2}m_{2}}(\bm{r_{2}})+\phi_{n_{2}l_{2}m_{2}}(\bm{r_{2}})\phi_{n_{1}l_{1}m_{1}}(\bm{r_{1}}) \right).
\label{eq:appNonproduct}
\end{equation}
\end{itemize}
 
For the wave-function to contribute to the first-order perturbation calculation then
\begin{equation}
\nonumber
\bra{\Psi_{0}^{(0)}}H'\ket{\Psi_{k\neq 0}^{(0)}}\neq0.
\end{equation}
We shall use this to determine the quantum numbers we need to consider and hence simplify the numerical calculation.

\subsection{Simplifying the integrals}
\subsubsection{Product states}
We first consider the $v_{\text{xc}}$ contribution from the product states $\phi_{n_{1}l_{1}m_{1}}(\bm{r_{1}})\phi_{n_{2}l_{2}m_{2}}(\bm{r_{2}})$.  Then $v_{\text{xc}}$ enters into the perturbation calculation as 

\begin{equation}
\nonumber
\bra{\phi_{000}(\bm{r_{1}})\phi_{000}(\bm{r_{2}})}\sum_{i=1,2}v_{\text{xc}}^{\text{LDA}}(r_{i})\ket{\phi_{n_{1}l_{1}m_{1}}(\bm{r_{1}})\phi_{n_{2}l_{2}m_{2}}(\bm{r_{2}})}
\end{equation} 
\begin{eqnarray}
\nonumber =\int \int R_{00}^{*}(r_{1})R_{00}^{*}(r_{2})\sum_{i=1,2}v_{xc}^{LDA}(r_{i})R_{n_{1}l_{1}}(r_{1})R_{n_{2}l_{2}}(r_{2})\\
\nonumber r_{1}^{2}r_{2}^{2}dr_{1}dr_{2}\int Y^{*}_{00}Y_{l_{1}m_{1}} d\Omega_{1}\int Y^{*}_{00}Y_{l_{2}m_{2}} d\Omega_{2}. 
\end{eqnarray} 

From the orthogonality of the spherical harmonics then $m_{1}=m_{2}=l_{1}=l_{2}=0$ for this to be a non-zero integral and from the orthogonality of the ground state radial solutions $n_{1}=n_{2}=0$ as well. Hence there are no contributions from the $v_{\text{xc}}$ term for excited states of this form.  This result also applies to the term arising from $v_{\text{H}}$.

Next we consider the Coulomb term, which may be expanded as a series of spherical harmonics:

\begin{eqnarray}
\nonumber&~&\frac{1}{|\bm{r}_{1}-\bm{r}_{2}|}=\sum_{l'=0}^{\infty}\sum_{m'=-l'}^{+l'}\frac{4\pi}{(2l'+1)}\frac{(r_{<})^{l'}}{(r_{>})^{l'+1}}\\
&~&Y_{l'm'}^{*}(\theta_{1},\phi_{1})Y_{l'm'}(\theta_{2},\phi_{2}),
\label{eqn:appCexpansion}
\end{eqnarray}

where $r_{<}=\min\{r_{1},r_{2}\}$ and $r_{>}=\max\{r_{1},r_{2}\}$.  We then use
\begin{equation}
Y^{*}_{lm}=(-1)^{m}Y_{l-m}
\label{eqn:spherid}
\end{equation}
in Eq.~(\ref{eqn:appCexpansion}) to employ the orthogonality of the spherical harmonics of the excited states to the series.  This, together with the required symmetry on exchange of particles, forcing $n_{1}=n_{2}$ and $l_{1}=l_{2}$, means the wave-function must be of the form $\phi_{nlm}(\bm{r_{1}})\phi_{nl-m}(\bm{r_{2}})$.

We have, again courtesy of orthogonality, that
\begin{equation}
\nonumber
\bra{\phi_{000}(\bm{r_{1}})\phi_{000}(\bm{r_{2}})}\frac{1}{|\bm{r}_{1}-\bm{r}_{2}|}\ket{\phi_{nlm}(\bm{r_{1}})\phi_{nl-m}(\bm{r_{2}})}
\end{equation}

\begin{eqnarray}
\nonumber&~&=\frac{1}{4\pi}(-1)^{m}\frac{4\pi}{(2l+1)}\int \int R_{00}^{*}(r_{1})R_{00}^{*}(r_{2})\frac{(r_{<})^{l}}{(r_{>})^{l+1}}\\
\nonumber&~&R_{nl}(r_{1})R_{nl}(r_{2})r_{1}^{2}r_{2}^{2}dr_{1}dr_{2}=\frac{1}{4\pi}(-1)^{m}\frac{4\pi}{(2l+1)}A,
\end{eqnarray}
where $A$ is understood to be the previous double integral and we have used Eq.~(\ref{eqn:spherid}) plus the explicit form of $Y_{00}$.

Now as the energy eigenvalues have no $m$ dependence, we are at liberty to sum over $m$ when calculating the contribution of an excited state.  Hence the contribution to Eq.~(\ref{eqn:appCexpansion}) from the product states of energy $E_{nl}$ will be

\begin{eqnarray}
\nonumber&~& S_{E_{nl}}= \frac{1}{E_{0}^{(0)}-E_{nl}^{(0)}}\frac{4\pi}{(2l+1)}\frac{A}{4\pi}\\
\nonumber&~&\sum_{m=-l}^{m=+l}(-1)^{m}R_{nl}(r_{1})R_{nl}(r_{2}) Y_{lm}(\theta_{1},\phi_{1})Y_{l-m}(\theta_{2},\phi_{2})
\end{eqnarray}
\begin{eqnarray}
\nonumber&~&=\frac{1}{E_{0}^{(0)}-E_{nl}^{(0)}}\frac{4\pi}{(2l+1)}\frac{A}{4\pi}\sum_{m=-l}^{m=+l}(-1)^{2m}R_{nl}(r_{1})R_{nl}(r_{2})\\
\nonumber&~& Y_{lm}(\theta_{1},\phi_{1})Y_{lm}^{*}(\theta_{2},\phi_{2}).
\end{eqnarray}
Here we have used Eq.~(\ref{eqn:spherid}).  Then using that
\begin{equation}
\nonumber
P_{l}(cos(\theta))=\frac{4\pi}{(2l+1)}\sum_{m=-l}^{m=+l}Y_{lm}(\theta_{1},\phi_{1})Y_{lm}^{*}(\theta_{2},\phi_{2}),
\end{equation}
where $\theta$ is the angle between $\bm{r_{1}}$ and $\bm{r_{2}}$, we obtain the contribution due to product states of energy $E_{nl}$ as 
\begin{equation}
\nonumber
S_{E_{nl}}=\frac{1}{E_{0}^{(0)}-E_{nl}^{(0)}}A\frac{1}{4\pi}R_{nl}(r_{1})R_{nl}(r_{2})P_{l}(cos(\theta)),
\end{equation}
which is symmetric on exchange of positions as required.

\subsubsection{Non-product states}

Let us consider for the non-product state (Eq~.\ref{eq:appNonproduct}) the contribution from the $v_{\text{xc}}$ term, when one electron is in the ground state.  

Let us define $\psi_{G}(\bm{r_{1}},\bm{r_{2}})=\phi_{000}(\bm{r_{1}})\phi_{000}(\bm{r_{2}})$ and 
$\psi_{E}(\bm{r_{1}},\bm{r_{2}})=\frac{1}{\sqrt{2}}\left(\phi_{000}(\bm{r_{1}})\phi_{nlm}(\bm{r_{2}})+\phi_{000}(\bm{r_{2}})\phi_{nlm}(\bm{r_{1}})\right)$ then

\begin{eqnarray}
\nonumber&~&\bra{\psi_{G}(\bm{r_{1}},\bm{r_{2}})}\sum_{i=1,2}v_{\text{xc}}(r_{i})\ket{\psi_{E}(\bm{r_{1}},\bm{r_{2}}}\\
\nonumber&~&=2\int\int \psi_{G} v_{\text{xc}}(r_{1}) \psi_{E}\bm{dr_{1}}\bm{dr_{2}},
\end{eqnarray}
where we have used the symmetry of the potential on exchanging $\bm{r_{1}}$ and $\bm{r_{2}}$.
We note that one of the single electron wave-functions must be in the ground state for these integrals to be non-zero. 
By orthogonality the previous equation becomes
\begin{eqnarray}
\nonumber&~&\bra{\psi_{G}}\sum_{i=1,2}v_{\text{xc}}(r_{i}) \ket{\psi_{E}}= \frac{2}{\sqrt{2}}\int\int\phi_{000}(\bm{r_{1}})\phi_{000}(\bm{r_{2}})\\
\nonumber&~& v_{\text{xc}}(r_{1})\phi_{000}(\bm{r_{2}})\phi_{nlm}(\bm{r_{1}})\bm{dr_{1}}\bm{dr_{2}}\\
\nonumber&~&=\frac{2}{\sqrt{2}}\int\int R_{00}(r_{1})v_{\text{xc}}(r_{1})R_{nl}(r_{1})r_{1}^{2}dr_{1}\int Y^{*}_{00}Y_{lm} d\Omega_{1},
\end{eqnarray}
where we have used the fact that the wave-function is normalised.  Hence $l=m=0$ for this contribution to be non-zero, with a similar result for $v_{\text{H}}$.

For the Coulomb term of this non-product state, the use of the expansion in spherical harmonics Eq.~(\ref{eqn:appCexpansion}) forces $l=m=0$ to ensure that the integral is non zero, with result
\begin{eqnarray}
\nonumber \bra{\psi_{G}} \frac{1}{|\bm{r}_{1}-\bm{r}_{2}|} \ket{\psi_{E}}= \frac{1}{\sqrt{2}}4\pi\int\int R_{00}(r_{1})R_{00}(r_{2})\frac{1}{r_{>}}\\
\nonumber \left(R_{00}(r_{1})R_{n0}(r_{2})+R_{00}(r_{2})R_{n0}(r_{1}) \right)r_{1}^{2}r_{2}^{2}dr_{1}dr_{2}\frac{1}{4\pi}.
\end{eqnarray}
Next we consider the non-product state where neither electron is in the ground state.  As neither single particle wave-function now has $n=0$ and $l=0$, then this means that there is no contribution from terms with $v_{\text{xc}}$ or $v_{\text{H}}$ for this form.  For there to be a Coulomb contribution, we have that from before $l_{1}=l_{2}$ and $m_{1}=-m_{2}$, so we consider only states of the following form
\begin{equation}
\nonumber
\psi_{E}=\frac{1}{\sqrt{2}}\left(\phi_{n_{1}lm}(\bm{r_{1}})\phi_{n_{2}l-m}(\bm{r_{2}})+\phi_{n_{1}lm}(\bm{r_{2}})\phi_{n_{2}l-m}(\bm{r_{1}})\right).
\end{equation}
A similar analysis to that employed in the previous sub-section allows the perturbation contribution from wave-functions with energy $E_{n_{1}l}+E_{n_{2}l}$ to be written in terms of the radial part and Legendre polynomials:
 
\begin{eqnarray}
\nonumber&~&S_{E_{nl}}=\frac{1}{E_{0}^{(0)}-E_{n_{1}l}^{(0)}-E_{n_{2}l}^{(0)}}\frac{B}{8\pi}\\
\nonumber&~&(R_{n_{1}l}(r_{1})R_{n_{2}l}(r_{2})+R_{n_{1}l}(r_{2})R_{n_{2}l}(r_{1}))P_{l}(cos(\theta)),
\end{eqnarray}
where
\begin{eqnarray}
\nonumber&~&B=\int \int R_{00}^{*}(r_{1})R_{00}^{*}(r_{2})\frac{(r_{<})^{l}}{(r_{>})^{l+1}}\\
\nonumber&~& \left (R_{n_{1}l}(r_{1})R_{n_{2}l}(r_{2})+R_{n_{1}l}(r_{2})R_{n_{2}l}(r_{1}) \right)r_{1}^{2}r_{2}^{2}dr_{1}dr_{2}.
\end{eqnarray}

\subsection{Calculating the density and reduced density matrix}
By grouping the wave-functions by $l$, we can write to first-order
\begin{eqnarray}
\nonumber&~&\Psi_{0}= f(r_{1},r_{2})P_{0}(\cos(\theta))+g(r_{1},r_{2})P_{1}(\cos(\theta))+\\
\nonumber&~&h(r_{1},r_{2})P_{2}(\cos(\theta))+\cdots
\end{eqnarray}
Using the orthogonality of the Legendre polynomials
\begin{equation}
\nonumber
\int P_{l}(\cos(\theta))P_{k}(\cos(\theta))d\Omega=\frac{4\pi}{2l+1}\delta_{lk}
\end{equation}
gives the easily computable expression for the density
\begin{equation}
\nonumber
n=2\int f^{2} dr_{1}4\pi+2\int g^{2} dr_{1}\frac{4\pi}{3}+2\int e^{2} dr_{1} \frac{4\pi}{5}+\cdots
\end{equation}
A similar treatment gives the reduced density matrix.


\begin{references}
\bibitem{NOTE1} By the Hohenberg Kohn theorem\cite{HK} the ground state density for an interacting electron system uniquely determines its ground state wave-function and hence all of the ground-state properties of the system including the entanglement.
\bibitem{NOTE2} The first method we tried was to vary the form of such a potential until the ground state of the corresponding Hamiltonian reproduces $n_{\text{LDA}}(r)$.  This search is difficult in the general case but is not too computationally intensive if we assume that the Hamiltonian is separable in centre of mass and relative motion co-ordinates, and solve the resulting radial equations numerically.  However, this then limits the search to $v_{\text{ext}}^{\text{LDA}}=\omega_{1}(r_{1}+r_{2})+\omega_{2}(r_{1}^{2}+r_{2}^{2})$ with $\omega_{1}$ and $\omega_{2}$ as parameters.  Due to this constraint this method was not successful in reproducing the density.
\end{references}
\end{document}